\documentclass[aps,prd,superscriptaddress,showpacs,preprintnumbers]{revtex4}
\usepackage{graphicx,color,amsmath}
\usepackage{textcomp}
\usepackage{cancel}
\usepackage{soul}
\newcommand{\be}{\begin{equation}}
\newcommand{\ee}{\end{equation}}
\newcommand{\bea}{\begin{eqnarray}}
\newcommand{\eea}{\end{eqnarray}}

\newcommand{\bfp}{\mbox{\boldmath $p$}}
\def\bpo{{\bfp}_{\perp 1}}
\def\bpt{{\bfp}_{\perp 2}}

\newcommand{\bfq}{\mbox{\boldmath $q$}}
\newcommand{\bfP}{\mbox{\boldmath $P$}}

\def\ppo{p_{\perp 1}}
\def\ppt{p_{\perp 2}}
\def\pp{p_\perp}

\newcommand{\qup}{q^\uparrow}

\newcommand{\ua}{\uparrow}

\def\avk{\langle k_\perp ^2\rangle}
\def\avp{\langle p_\perp ^2\rangle}
\def\avk{\langle k_\perp ^2\rangle}
\def\avp{\langle p_\perp ^2\rangle}

\def\C{_{_C}}

\newcommand{\deltatilde}{\tilde{\Delta} ^N D}
\newcommand{\yesp}{{\color{red} \Large \textbullet}}
\newcommand{\nop}{{\color{blue} \Large \textopenbullet}}

\def\lsim{\mathrel{\rlap{\lower4pt\hbox{\hskip1pt$\sim$}}\raise1pt\hbox{$<$}}}
\def\gsim{\mathrel{\rlap{\lower4pt\hbox{\hskip1pt$\sim$}}\raise1pt\hbox{$>$}}}
\def\nostrocostruttino#1\over#2{\mathrel{\mathop{\kern 0pt \rlap
{\hbox{$#1$}}} \hbox{\kern-.135em $#2$}}}

\begin{document}

\title{Extracting the Kaon Collins function from $e^+e^-$ hadron pair production
data}

\author{M.~Anselmino}
\affiliation{Dipartimento di Fisica, Universit\`a di Torino,
             Via P.~Giuria 1, I-10125 Torino, Italy}
\affiliation{INFN, Sezione di Torino, Via P.~Giuria 1, I-10125 Torino, Italy}
\author{M.~Boglione}
\affiliation{Dipartimento di Fisica, Universit\`a di Torino,
             Via P.~Giuria 1, I-10125 Torino, Italy}
\affiliation{INFN, Sezione di Torino, Via P.~Giuria 1, I-10125 Torino, Italy}
\author{U.~D'Alesio}
\affiliation{Dipartimento di Fisica, Universit\`a di Cagliari,
             I-09042 Monserrato (CA), Italy}
\affiliation{INFN, Sezione di Cagliari,
             C.P.~170, I-09042 Monserrato (CA), Italy}
\author{J.O.~Gonzalez~Hernandez}
\affiliation{Dipartimento di Fisica, Universit\`a di Torino,
             Via P.~Giuria 1, I-10125 Torino, Italy}
\affiliation{INFN, Sezione di Torino, Via P.~Giuria 1, I-10125 Torino, Italy}
\author{S.~Melis}
\affiliation{Dipartimento di Fisica, Universit\`a di Torino,
             Via P.~Giuria 1, I-10125 Torino, Italy}
\author{F.~Murgia}
\affiliation{INFN, Sezione di Cagliari,
             C.P.~170, I-09042 Monserrato (CA), Italy}
\author{A.~Prokudin}
\affiliation{Division of Science, Penn State Berks, Reading, PA 19610, USA}
\date{\today}

\begin{abstract}
The latest data released by the BaBar Collaboration on azimuthal correlations
measured for pion-kaon and kaon-kaon pairs produced in $e^+e^-$ annihilations
allow, for the first time, a direct extraction of the kaon Collins functions.
These functions are then used to compute the kaon Collins asymmetries in 
Semi Inclusive Deep Inelastic Scattering processes, which result in good 
agreement with the measurements performed by the HERMES and COMPASS Collaborations.
\end{abstract}



\maketitle

\section{\label{Intro} Introduction}

In the quest for the understanding of the inner 3D structure of nucleons,
the transverse momentum dependent partonic distribution and fragmentation
functions (respectively TMD-PDFs and TMD-FFs) play a fundamental role. In
particular, it is inside the TMD-FFs that we encode the non-perturbative,
soft part of the hadronisation process.

Over the years, combined analyses of Semi Inclusive Deep Inelastic
Scattering (SIDIS) and $e^+e^-\to \pi^+\pi^-X$
experimental data allowed the extraction of the transversity distribution
and the $\qup \to \pi\,X$ (pion) Collins functions~\cite{Anselmino:2007fs,
Anselmino:2008jk,Anselmino:2013vqa,Anselmino:2015sxa}.
However, until very recently, no direct experimental information
was available on the {\it kaon} Collins functions, although their effects
were clearly evident in SIDIS
processes~\cite{Airapetian:2010ds,Airapetian:2012yg,Alekseev:2008aa,Adolph:2014zba},
both in the $\cos 2\phi_h$ azimuthal modulation of the unpolarised cross
section and in the $\sin(\phi_h+\phi_S)$ azimuthal asymmetry, the so-called Collins asymmetry.

The Collins function, in fact, contributes to the $\cos 2\phi_h$ asymmetries
in convolution with a Boer-Mulders function, while in the $\sin(\phi_h+\phi_S)$
single spin  asymmetries it appears convoluted with the transversity distribution.
The kaon $\cos 2\phi_h$ azimuthal asymmetries present some peculiar features:
at HERMES~\cite{Airapetian:2012yg} $K^+$ and $K^-$
asymmetries are both
sizeable and negative, while the analogous $\pi^+$ asymmetries are compatible
with zero or slightly negative and the $\pi^-$ ones are positive.
Looking at the $\sin(\phi_h+\phi_S)$ dependence, instead, we observe that $K^+$ asymmetries look slightly positive, 
while $K^-$ data are compatible with zero (within large errors)~\cite{Airapetian:2010ds,Adolph:2014zba}.

Clearly, to understand better these data we have to study the kaon Collins
functions. Recent BaBar data on pion-pion, pion-kaon and kaon-kaon production
from $e^+e^-$ annihilation processes~\cite{Aubert:2015hha} give the opportunity to extract
the kaon Collins function, for the first time; moreover, all these results
have been presented in the same bins of $z_1$ and $z_2$, so that they can be
analysed simultaneously in a consistent way.

In this paper we perform an analysis of the $e^+e^-$ BaBar measurements
involving kaons, with the aim of extracting the kaon Collins functions.
This paper extends a recent study of the Collins functions in $e^+e^-$
and SIDIS data~\cite{Anselmino:2015sxa} limited to pion production.
Our strategy is the following:
\begin{enumerate}
 \item
 When necessary for our analysis (for instance for the description of  
 $e^+e^- \to \pi K X$ data) we employ
 the favoured and disfavoured pion Collins functions obtained in
 Ref.~\cite{Anselmino:2015sxa}: no free parameters are introduced in this
 analysis concerning pions.
 \item
 We parameterise the kaon favoured and disfavoured Collins functions using
 a factorised form, similar to that used for pions~\cite{Anselmino:2015sxa},
 with an even simpler structure: due to the limitation of the kaon data
 presently available, we have found out, after several tests,
 that it suffices for their analysis to consider a model which implies only
 two free parameters, instead of four. We also do not introduce different
 parameters between heavy and light flavours in the kaon Collins functions
 (this point will be further discussed at the end of Section~\ref{Fit}).
 The free parameters will be determined by best fitting the new $e^+e^-\to
 \pi \, K \, X$ and $e^+e^-\to K^+ \, K^- \, X$  BaBar data
 sets~\cite{Aubert:2015hha}.
 \item
 The kaon favoured and disfavoured Collins functions extracted from $e^+e^-$
 annihilation data will be used to compute the values of the 
  Collins single spin 
 asymmetries observed in SIDIS processes. As we will discuss in
 Section~\ref{Fit}, the comparison of our
 predictions with the measurements performed by the HERMES and COMPASS
 Collaborations confirms, within the precision limits of experimental data,
 the total consistency of the Collins functions extracted from $e^+e^-$ data
 with those 
 obtained from SIDIS processes, corroborating their  
 universality~\cite{Collins:2004nx}.

 \end{enumerate}

In Section~\ref{Form} we briefly recall the formalism used in our analysis,
while in Section~\ref{Fit} we present the results of our best fits of BaBar
kaon data and compare them with SIDIS measurements of the kaon Collins
asymmetry. Some short final comments and conclusions will be given in
Section~\ref{Com}.

\section{\label{Form} Formalism}


In this section we briefly summarise the formalism relevant to perform the
extraction of the kaon Collins functions using the new data from the BaBar
Collaboration, which now contain also asymmetries for $e^+e^-$ annihilations
into pion-kaon and kaon-kaon pairs. Two methods have been adopted in the
experimental analysis, the so called ``thrust-axis method'' and the
``hadronic plane method''. Here, we concentrate on the latter and refer
the reader to our previous simultaneous analyses of SIDIS and
$e^+e^-\to \pi \, \pi \, X$ data~\cite{Anselmino:2015sxa} for further details. 

\subsection{Parameterisation of the kaon Collins function}

For the unpolarised parton distribution and fragmentation
functions we adopt a simple factorised form, in which longitudinal and
transverse degrees of freedom are separated. The dependence on the intrinsic
transverse momentum is assumed to have a Gaussian shape:
\bea
 f_{q/p}(x,k_\perp) & = &
 f_{q/p}(x)\;\frac{e^{-{k_\perp^2}/\avk}}{\pi\avk} \label{funp} \\
 D_{h/q}(z,\pp)& =& D_{h/q}(z)\;\frac{e^{-\pp^2/\avp}}{\pi\avp}
 \label{dunp}\;,
\eea
with $\langle k_\perp^2\rangle =0.57 \; {\rm GeV}^2$ and
$\langle p_\perp^2\rangle =0.12 \; {\rm GeV}^2$ as found in
Ref.~\cite{Anselmino:2013lza} by analysing the HERMES unpolarised
SIDIS multiplicities.
For the collinear parton distribution and fragmentation functions,
$f_{q/p}(x)$ and $D_{h/q}(z)$, we use the GRV98LO PDF set~\cite{Gluck:1998xa}
and the DSS fragmentation function set from Ref.~\cite{deFlorian:2007hc}.

{F}or the Collins FF, $\Delta^N\! D_{h/q^\uparrow}(z,\pp)$, we adopt the
following parameterisation~\cite{Anselmino:2015sxa}:
\be
\Delta^N \! D_{h/q^\uparrow}(z,\pp) =  \tilde{\Delta} ^N D_{h/q^\ua}(z)
\> h(\pp)\,\frac{e^{-\pp^2/{\avp}}}{\pi \avp}\,,
\label{coll-funct}
\ee
where
\be
\tilde{\Delta} ^N D_{h/q^\ua}(z) = 2 \, {\cal N}^{\C}_{q}(z)\,
D_{h/q}(z)
\label{coll-D}
\ee
represents the $z$-dependent part of the Collins function at the initial scale $Q^2_0$,
which is then evolved to the appropriate value of $Q^2=112$ GeV$^2$.
In this analysis, we use a simple model which implies no $Q^2$ dependence in the
$\pp$ distribution. As the Collins function in our parameterisation is proportional
to the unpolarised fragmentation function, see Eq.~\eqref{coll-funct}
and~\eqref{coll-D}, we assume that the only scale dependence is contained in
$D(z,Q^2)$, which is evolved with an unpolarised DGLAP kernel, while
${\cal N}^{\C}_{q}$ does not evolve in $Q^2$. This amounts to assuming
that the ratio $\Delta^N D(z,p_{\perp},Q^2)/D(z,Q^2)$ is constant in $Q^2$.

The function $h(\pp)$, defined as
\be
 h(\pp)=\sqrt{2e}\,\frac{p_\perp}{M_{C}}\,e^{-{p_\perp^2}/{M_{C}^2}}
 \label{hpcollins}\,,
\ee
allows for a possible modification of the $\pp$ Gaussian width of the
Collins function with respect to the unpolarised FF, while fulfilling 
the appropriate positivity bound: this modification is
controlled by the parameter $M_C^2$.

{F}or the pion ${\cal N}^{\C}_{q}(z)$, we fix the favoured and disfavoured contributions
as obtained from the reference fit of Ref.~\cite{Anselmino:2015sxa}:
\bea
\mathcal{N}^{\C}_{\rm fav}(z) &=& N^{\pi}_{\rm fav} \,z^{\gamma}(1-z)^\delta\,\,
\frac{(\gamma+\delta)^{\gamma+\delta}}{\gamma^\gamma \delta^\delta}
\label{std-fav-pi} \\
\mathcal{N}_{\rm dis}^{\C}(z)&=& N^{\pi}_{\rm dis}\,,
\label{std-disf-pi}
\eea
with $N^{\pi}_{\rm fav}=0.90$, $N^{\pi}_{\rm dis}=-0.37$, $\gamma=2.02$ and $
\delta=0.00$, as reported in Table~\ref{fitpar-pi}.

{F}or the kaon we parameterise the favoured and disfavoured Collins contributions
by setting ${\mathcal N}^{\C}_{q}(z)$ to a constant:
\bea
\mathcal{N}^{\C}_{\rm fav}(z) &=& N^{K}_{\rm fav}
\label{std-fav-K} \\
\mathcal{N}_{\rm dis}^{\C}(z)&=&N^{K}_{\rm dis}\,,
\label{std-disf-K}
\eea
which brings us to a total of two free parameters for the Collins functions.
In fact, the experimental data presently available for kaon production do not
require a four-parameter fit, as in the pion case. We have indeed explicitly
checked that a four-parameter fit does not result in a lower value of the total
$\chi^2$.

\subsection{$e^+e^-\to h_1 \, h_2 \, X$ in the hadronic-plane method}

In the ``hadronic-plane method'' one adopts a reference frame in which one 
of the produced hadrons ($h_2$ in
our case) identifies the $\hat z$ direction and the $\widehat{xz}$ plane
is determined by the lepton and the $h_2$ directions; the other relevant
plane is determined by $\hat z$ and the direction of the other
observed hadron, $h_1$, at an angle $\phi_1$ with respect to the
$\widehat{xz}$ plane; $\theta_2$ is the angle between $h_2$ and the $e^+e^-$ direction.

In this case, the elementary process $e^+e^- \to q \, \bar q$
does not occur in the $\widehat{xz}$ plane, and thus the helicity scattering
amplitudes involve an azimuthal phase, $\varphi_2$.
The differential cross section reads
\bea
\frac{d\sigma ^{e^+e^-\to h_1 h_2 X}}
{dz_1\,dz_2\,d^2\bpo\,d^2\bpt\,d\cos\theta_2}&=&
 \frac{3\pi\alpha^2}{2s} \, \sum _q e_q^2 \, \Big\{
 (1+\cos^2\theta_2)\,D_{h_1/q}(z_1,\ppo)\,D_{h_2/\bar q}(z_2,\ppt)
\frac{}{} \Big.  \!\!\! \label{belle2} \\ & &
+ \Big. \frac{1}{4}\,\sin^2\theta_2\,\Delta ^N D _{h_1/q^\ua}(z_1,\ppo)\,
 \Delta ^N D _{h_2/\bar q^\ua}(z_2,\ppt)\,
\cos(2\varphi_2 + \phi_{q}^{h{_1}})\Big\}\,,\nonumber
\eea
where $\phi_{q}^{h_1}$ is the azimuthal angle of the detected hadron $h_1$
around the direction of the parent fragmenting quark, $q$. In other words,
$\phi_{q}^{h_1}$ is the azimuthal angle of $\bpo$ in the helicity frame of
$q$. It can be expressed in terms of
$\bpt$ and $\bfP_{1T}$, the transverse momentum of the $h_1$ hadron in the hadronic-plane reference frame.
At lowest order in $\pp/(z\sqrt{s})$ we have
\bea
&&\cos\phi_{q}^{h_1} = \frac{P_{1T}}{\ppo} \,
\cos(\phi_1-\varphi_2) - \frac{z_1}{z_2} \, \frac{\ppt}{\ppo} \\
&&\sin\phi_{q}^{h_1}=
\frac{P_{1T}}{\ppo} \, \sin(\phi_1-\varphi_2) \;.
\eea
Using the parameterisation of the Collins function given in Eqs.~\eqref{coll-funct}-\eqref{hpcollins},
the integration over $\bpt$ in Eq.~\eqref{belle2} can be performed explicitly.
Moreover, since $\bpo = \bfP_{1} -z_1\bfq _1$, we can replace $d^2\bpo$ with $d^2\bfP _{1T}$.
Integrating also over $P_{1T}$, but not over $\phi_1$, we then obtain
\be
\frac{d\sigma ^{e^+e^-\to h_1 h_2 X}}
{dz_1\,dz_2\,d\cos\theta_2\,d\phi_1} =
\frac{3\alpha^2}{4s}\,\left\{\, D^{h_1 h_2} + N^{h_1 h_2}\,\cos(2\phi_1)\,\right\}\,,
\label{belle3}
\ee
where
\bea
D^{h_1 h_2} &=& (1+\cos^2\theta_2)\,\sum _q e_q^2\, D_{h_1/q}(z_1)\,
D_{h_2/\bar q}(z_2)\label{Dhh} \,\\
N^{h_1 h_2} &=&\frac{1}{4}\,\frac{z_1z_2}{z_1^2+z_2^2}\,\sin^2\theta_2\,
\frac{2e\,\avp M_C^4}{(\avp+M_C^2)^3}\,
\sum_q e^2_q \, \deltatilde_{h_1/q^\ua}(z_1)\,
\deltatilde_{h_2/\bar q^\ua}(z_2)\,.
\label{Nhh}
\eea

By normalising this result to the azimuthal averaged cross section
\be
\langle\,d\sigma\,\rangle = \frac{1}{2\pi}\,
\frac{d\sigma ^{e^+e^-\to h_1 h_2 X}}{dz_1\,dz_2\,d\cos\theta_2} =
\frac{3\alpha^2}{4s}\,D^{h_1 h_2}\,,
\label{csaver}
\ee
one gets
\be
R_{0}^{h_1h_2} \equiv \frac{1}{\langle\,d\sigma\,\rangle} \>
\frac{d\sigma ^{e^+e^-\to h_1 h_2 X}}
{dz_1\,dz_2\,d\cos\theta_2\,d\phi_1}
= 1+P_{0}^{h_1h_2} \, \cos(2\phi_1)\,,
\label{A0h1h2}
\ee
having defined
\be
P_{0}^{h_1h_2} = \frac{N^{h_1h_2}}{D^{h_1h_2}}\,\cdot
\ee

In our previous analysis~\cite{Anselmino:2015sxa}, we considered the like sign (L),
unlike sign (U) and charged (C) combinations for pion-pion pairs, which are constructed by
using the appropriate combinations of charged pions, that is, by replacing
$P_{0}^{h_1h_2}$ in Eq.~\eqref{A0h1h2} by 
\begin{subequations}
\bea
P_{0L}^{\pi\pi} \equiv \frac{N_{L}^{\pi\pi}}{D_{L}^{\pi\pi}} &=&
\frac{N^{\pi^+\pi^+} + N^{\pi^-\pi^-}}{D^{\pi^+\pi^+} + D^{\pi^-\pi^-}}  \\
P_{0U}^{\pi\pi} \equiv \frac{N_{U}^{\pi\pi}}{D_{U}^{\pi\pi}} &=&
\frac{N^{\pi^+\pi^-} + N^{\pi^-\pi^+}}{D^{\pi^+\pi^-} + D^{\pi^-\pi^+}}  \\
P_{0C}^{\pi\pi} \equiv \frac{N_{C}^{\pi\pi}}{D_{C}^{\pi\pi}} &=&
\frac{N_{L}^{\pi\pi} + N_{U}^{\pi\pi}}{D_{L}^{\pi\pi} + D_{U}^{\pi\pi}}  \; \cdot
\eea
\label{P0p}
\end{subequations}
Analogously, for kaon-kaon pairs:
\begin{subequations}
\bea
P_{0L}^{KK} \equiv \frac{N_{L}^{KK}}{D_{L}^{KK}} &=&
\frac{N^{K^+K^+} + N^{K^-K^-}}{D^{K^+K^+} + D^{K^-K^-}}  \\
P_{0U}^{KK} \equiv \frac{N_{U}^{KK}}{D_{U}^{KK}} &=&
\frac{N^{K^+K^-} + N^{K^-K^+}}{D^{K^+K^-} + D^{K^-K^+}}  \\
P_{0C}^{KK} \equiv \frac{N_{C}^{KK}}{D_{C}^{KK}} &=&
\frac{N_{L}^{KK} + N_{U}^{KK}}{D_{L}^{KK} + D_{U}^{KK}    }  \; ,
\eea
\label{P0k}
\end{subequations}
and for pion-kaon production:
\begin{subequations}
\bea
P_{0L}^{\pi K} \equiv \frac{N_{L}^{\pi K}}{D_{L}^{\pi K}} &=&
\frac{N^{\pi^+K^+} + N^{\pi^-K^-} + N^{K^+\pi^+} + N^{K^-\pi^-}}{D^{\pi^+K^+} +
D^{\pi^-K^-} + D^{K^+\pi^+} + D^{K^-\pi^-}}  \\
P_{0U}^{\pi K} \equiv \frac{N_{U}^{\pi K}}{D_{U}^{\pi K}} &=&
\frac{N^{\pi^+K^-} + N^{\pi^-K^+} + N^{K^+\pi^-} + N^{K^-\pi^+}}{D^{\pi^+K^-} +
D^{\pi^-K^+} + D^{K^+\pi^-} + D^{K^-\pi^+}}  \\
P_{0C}^{\pi K} \equiv \frac{N_{C}^{\pi K}}{D_{C}^{\pi K}} &=&
\frac{N_{L}^{\pi K} + N_{U}^{\pi K}}{D_{L}^{\pi K} + D_{U}^{\pi K}}  \; \cdot
\eea
\label{P0pk}
\end{subequations}
We can now build ratios of unlike/like and unlike/charged asymmetries:
\be
\frac{(R_0^{h_1h_2})^{U}}{(R_0^{h_1h_2})^{L(C)}}= \frac{1+P_{0U}^{h_1h_2}
\, \cos(2\phi_1)}{1+P_{0L(C)}^{h_1h_2}  \, \cos(2\phi_1)}
\simeq 1+(P_{0U}^{h_1h_2} - P_{0L(C)}^{h_1h_2})  \, \cos(2\phi_1)
\label{A0ulc}\,,
\ee
where $P_{0U}^{h_1h_2}$, $P_{0L}^{h_1h_2}$ and $P_{0C}^{h_1h_2}$ can be taken
from Eqs.~\eqref{P0p}-\eqref{P0pk}.
Finally, one can write the asymmetries that are measured experimentally,
which correspond to the coefficient of the cosine in Eq.~\eqref{A0ulc}:
\bea
(A_0^{h_1h_2})^{UL} &=& P_{0U}^{h_1h_2} - P_{0L}^{h_1h_2} \\
(A_0^{h_1h_2})^{UC} &=& P_{0U}^{h_1h_2} - P_{0C}^{h_1h_2} .
\eea
%

\section{\label{Fit} Best fitting and results}

As mentioned above, we have adopted the following procedure:
\begin{enumerate}
 \item
 We employ the pion favoured and disfavoured  Collins functions as obtained
 in our recent extraction~\cite{Anselmino:2015sxa} based on 
 BaBar~\cite{TheBABAR:2013yha} and 
 Belle~\cite{Seidl:2008xc,Seidl:2012er} 
 $e^+e^- \to \pi\,\pi \, X$ data. As far as pions are concerned no free
 parameters are introduced in this analysis.
 The fixed values of the pion Collins function parameters are presented
 in Table~\ref{fitpar-pi}, together with the parameters obtained for the transversity
 distribution, which are given for later use.
%
%
\begin{table}[ht]
\renewcommand{\tabcolsep}{0.4pc} 
\renewcommand{\arraystretch}{1.5} 

\begin{tabular}{|l|l|}
\hline
 $ N_{u_v}^{T} =0.61^{+0.39}_{-0.23} $ & $N_{d_v}^{T} =-1.00^{+1.86}_{-0.00}$ \\
 $ \alpha =0.70^{+1.31}_{-0.63} $ & $\beta =1.80^{+7.60}_{-1.80}$ \\
\hline
 $ N_{\rm fav}^{\pi} =0.90^{+0.09}_{-0.34} $ & $N_{\rm dis}^{\pi} =-0.37^{+0.05}_{-0.05}$ \\
 $ \gamma =2.02^{+0.83}_{-0.33} $ & $\delta =0.00^{+0.42}_{-0.00}$ \\
 $ M^2_C  =0.28^{+0.20}_{-0.09} \textrm{ GeV}^2 $ &  \\
\hline
\end{tabular}
\caption{
Fixed parameters for the $u$ and $d$ valence quark
transversity distribution functions and the favoured and
disfavoured pion-Collins fragmentation functions, as obtained by fitting
simultaneously SIDIS data on the Collins asymmetry and Belle and BaBar data on
$A_{0}^{UL}$ and $A_{0}^{UC}$, for pion-pion pair production, in Ref.~\cite{Anselmino:2015sxa}.}
\label{fitpar-pi}
\end{table}
%
 \item
 The kaon favoured and disfavoured Collins functions are parameterized
 using a factorised form similar to that used for pions, but with a
 simpler structure: due to the limitations of the kaon data presently
 available, we introduce only two free parameters in our fit, instead of
 four, in such a way that the $z$-dependent part of the Collins functions
 will simply be proportional to their unpolarised counterparts:
\be
\tilde{\Delta} ^N D_{K/q^\ua}(z) = 2 \, N^{K}_{i}\,D_{K/q}(z) \,, \qquad i={\rm fav,dis}\,.
\label{coll-D-K}
\ee
$N^{K}_{\rm fav}$ and $N^{K}_{\rm dis}$ are free parameters to be fixed by
best fitting the experimental data. In this fit, which we denote as
our ``reference fit'', we make no distinction, for the values of $N^{K}_{i}$,
between heavy and light flavours; notice, however, that the favoured kaon
Collins functions for the $s$ quark will, in fact, be different from that
of the $u$ flavour: this difference is induced by the unpolarised, collinear
FFs used in our parameterisation, which imply consistently different
contributions for heavy and light flavours.
The Gaussian width of the kaon Collins function, controlled by the
parameter $M_C^2$, Eq.~\eqref{hpcollins}, is assumed to be the same as that
of the pion Collins function. Present data are not sensitive enough to the
shape of the $\pp$ dependence of the Collins functions to make further
distinctions. Moreover, for the same reason, no $Q^2$ dependence of the
$\pp$ distribution is included in our model. Further considerations on the
choice of two parameters will be made at the end of this Section.

This reference best fit gives the following results for the two free parameters considered:
\be
N_{\rm fav}^{K} =0.41^{+0.10}_{-0.10}\,, \qquad \qquad N_{\rm dis}^{K} =
0.08^{+0.18}_{-0.26}\,,
\label{fitpar-K}
\ee
suggesting a solution with a {\em positive} favoured Collins function, and a disfavoured contribution compatible 
with zero, within large errors. 
However, as we will discuss in Section~\ref{more-param}, a definite conclusion 
can only be drawn about the positive sign of the favoured light flavour 
contribution. Note that the pion Collins fragmentation 
functions extracted in Ref.~\cite{Anselmino:2015sxa} have opposite signs 
for favoured and disfavoured functions, and disfavoured functions are 
definitely non zero.

The contributions  to the total $\chi^2$ of each fitted set of data are
given in Table~\ref{chisq}. It is a good fit and, as one can see from
Figs.~\ref{fig:an-babar-piK} and~\ref{fig:an-babar-KK}, the data are described
well. The $A_0^{UL}$ asymmetries for $KK$ production 
are quite scattered and do not show a definite trend: it is for these data
that we obtain the largest $\chi^2$ contribution. The bands shown in
Figs.~\ref{fig:an-babar-piK} and~\ref{fig:an-babar-KK} are obtained by
sampling 1500 sets of parameters corresponding to a $\chi^2$ value in the
range between $\chi^2_{\rm min}$ and $\chi^2_{\rm min} + \Delta \chi^2 $,
as explained in Ref.~\cite{Anselmino:2015sxa}. The value of $\Delta \chi^2$
corresponds to 95.45\% confidence level for 2 parameters;
in this case we have $\Delta\chi^2=6.18$.
%
%
%
\begin{table}[ht]
\setlength{\tabcolsep}{10pt} 
\renewcommand{\arraystretch}{1.5} 
\begin{center}\setlength{\arraycolsep}{50pt}
\begin{tabular}{|c|c|c|c|}
\hline
 Data set    &$\chi^2$ &   points & $\chi^2/{\rm points}$ \\
\hline
$K\pi$ production \quad $A_0^{UL} $&  $14.6$ & $16$ & $0.91$\\
$K\pi$ production \quad $A_0^{UC} $&  $ 7.4$ & $16$ & $0.46$\\
$KK$   production   \quad $A_0^{UL} $&  $23.6$ & $16$ & $1.48$\\
$KK$   production   \quad $A_0^{UC} $&  $ 9.4$ & $16$ & $0.59$\\
\hline
\hline
Total  & 55.0 & 64 & $\chi^2_{\rm d.o.f.}=0.89$ \\
\hline
\end{tabular}
\end{center}
\caption{$\chi ^2$ values obtained in our reference fit. See text for details.}
\label{chisq}
\end{table}
%
%
%
\begin{figure}[h!t]
\includegraphics[width=8.6truecm,angle=0]{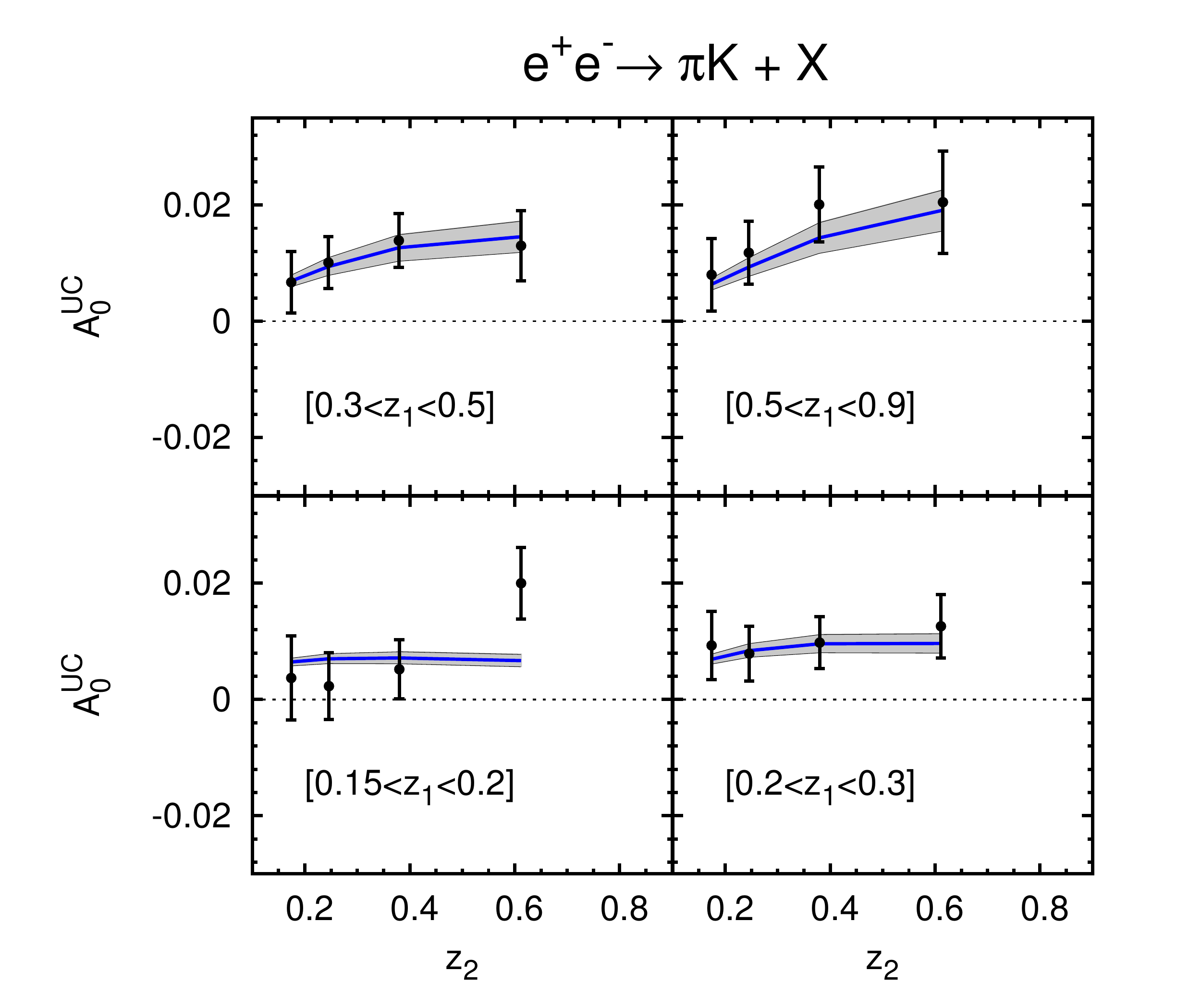}
\includegraphics[width=8.6truecm,angle=0]{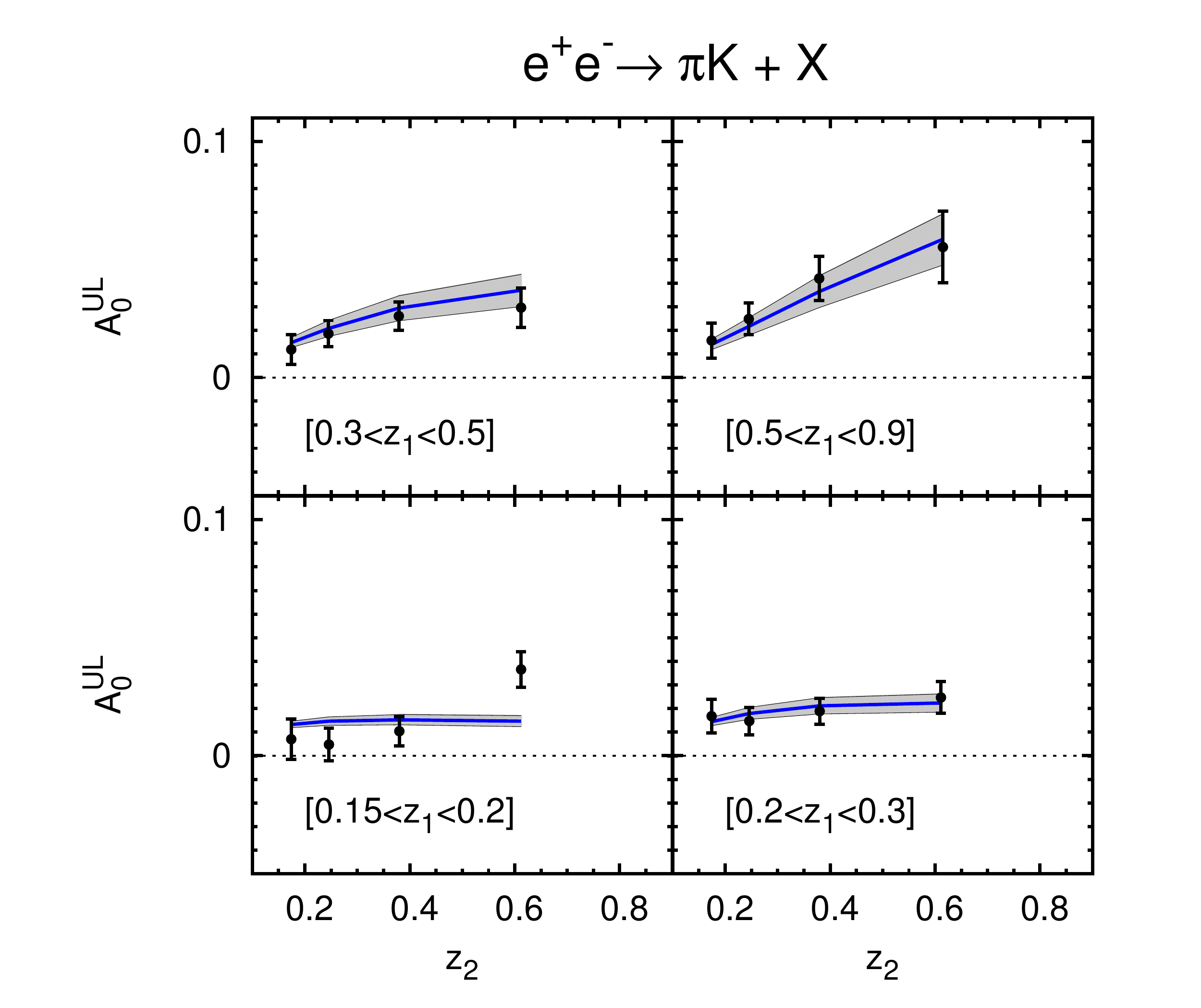}
\caption{The experimental data on the azimuthal correlations
$A_{0}^{UC}$ and $A_{0}^{UL}$  as functions of $z_1$ and $z_2$  in unpolarised
$e^+e^- \to \pi \,K\, X$ processes, as measured by the BaBar Collaboration,
are compared to the curves obtained from our reference fit, given by the 
parameters shown in Eq.~\eqref{fitpar-K}.
The shaded area corresponds to the statistical
uncertainty on these parameters.}
\label{fig:an-babar-piK}
\end{figure}
%
\begin{figure}[h!t]
\includegraphics[width=8.6truecm,angle=0]{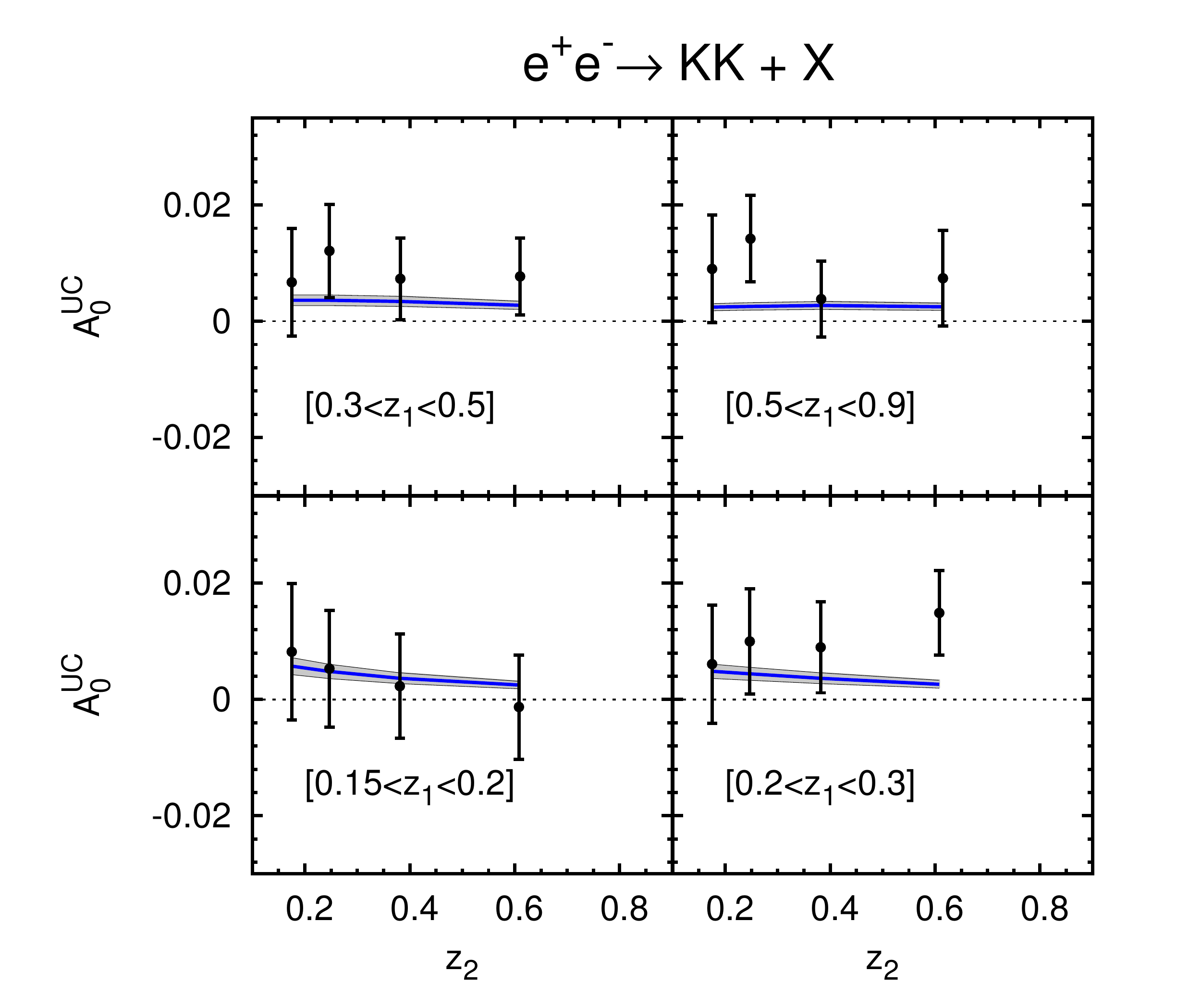}
\includegraphics[width=8.6truecm,angle=0]{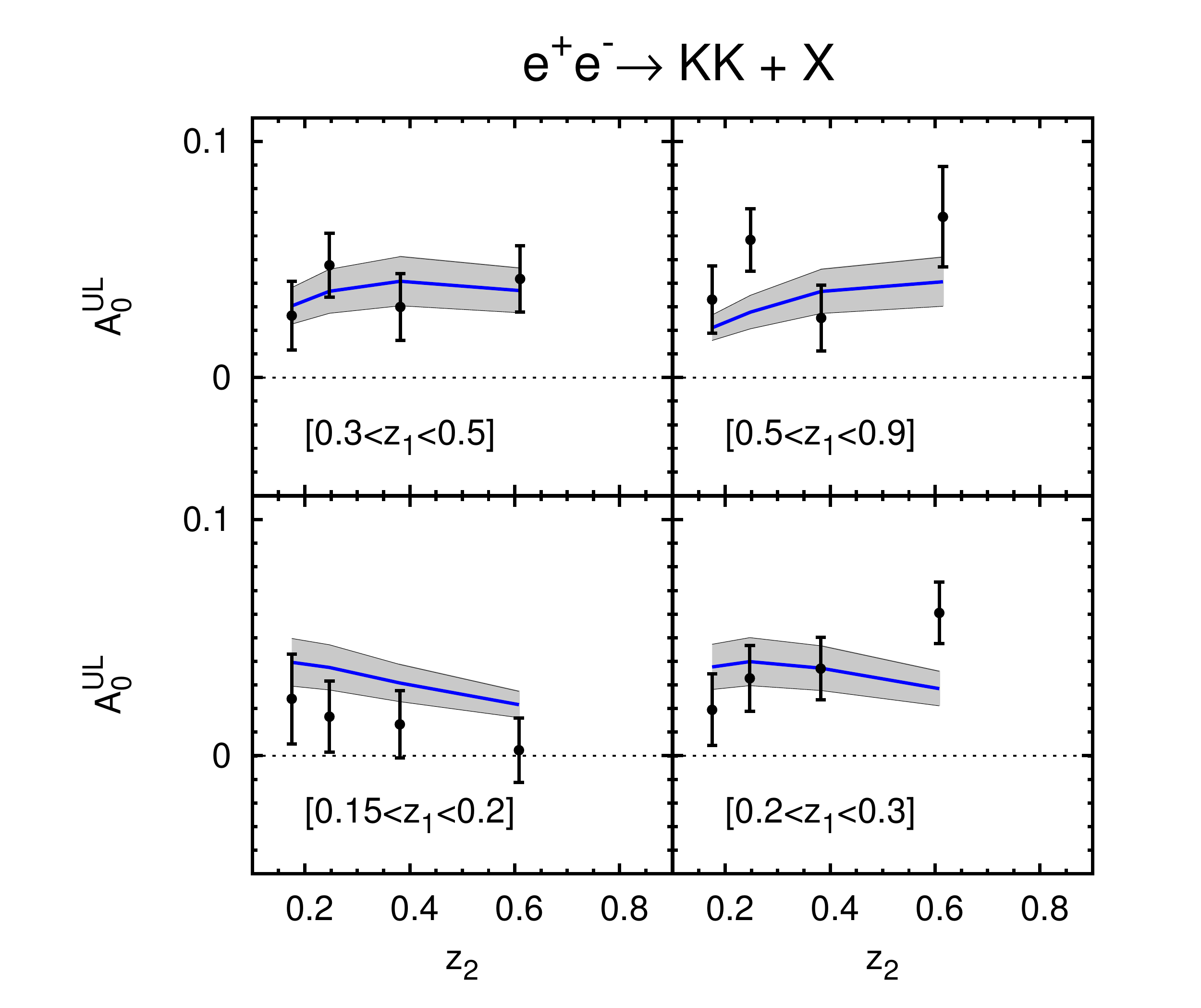}
\caption{The experimental data on the azimuthal correlations
$A_{0}^{UC}$ and $A_{0}^{UL}$  as functions of $z_1$ and $z_2$  in unpolarised
$e^+e^- \to K^+ K^- \, X$ processes, as measured by the BaBar Collaboration,
are compared to the curves obtained from our reference fit, given by the
parameters shown in Eq.~\eqref{fitpar-K}.
The shaded area corresponds to the statistical
uncertainty on these parameters.}
\label{fig:an-babar-KK}
\end{figure}
%

\item We deliberately choose not to include SIDIS kaon data in the fit at this stage. 
Including them would, in principle, require a 
global analysis of both pion and kaon data sets which is beyond the scope of this paper. 
Moreover, we would like to test the universality
of the Collins fragmentation functions in $e^+e^-$ and SIDIS, as proposed in 
Ref.~\cite{Collins:2004nx}, and  
check whether the kaon favoured and disfavoured
Collins functions extracted from $e^+e^-$ annihilation data can describe
the 
Collins asymmetries observed in SIDIS processes.
We 
compute the Collins SIDIS asymmetry
$A_{UT}^{\sin(\phi_h+\phi_S)}$, using the kaon Collins functions given by our
reference fit, Eqs.~\eqref{std-fav-K}, \eqref{std-disf-K} and~\eqref{fitpar-K},
and the transversity distributions obtained in Ref.~\cite{Anselmino:2015sxa}
and given in Table~\ref{fitpar-pi}.
The comparison of our predictions with the
measurements performed by the HERMES and COMPASS Collaborations is shown
in Figs.~\ref{fig:an-hermes-k} and \ref{fig:an-compass-k} respectively.
The good agreement confirms, within the precision limits of experimental
data, the consistency of the Collins functions extracted from $e^+e^-$
data with those active in SIDIS processes.
\begin{figure}[ht]
\includegraphics[width=10.5truecm,angle=0]{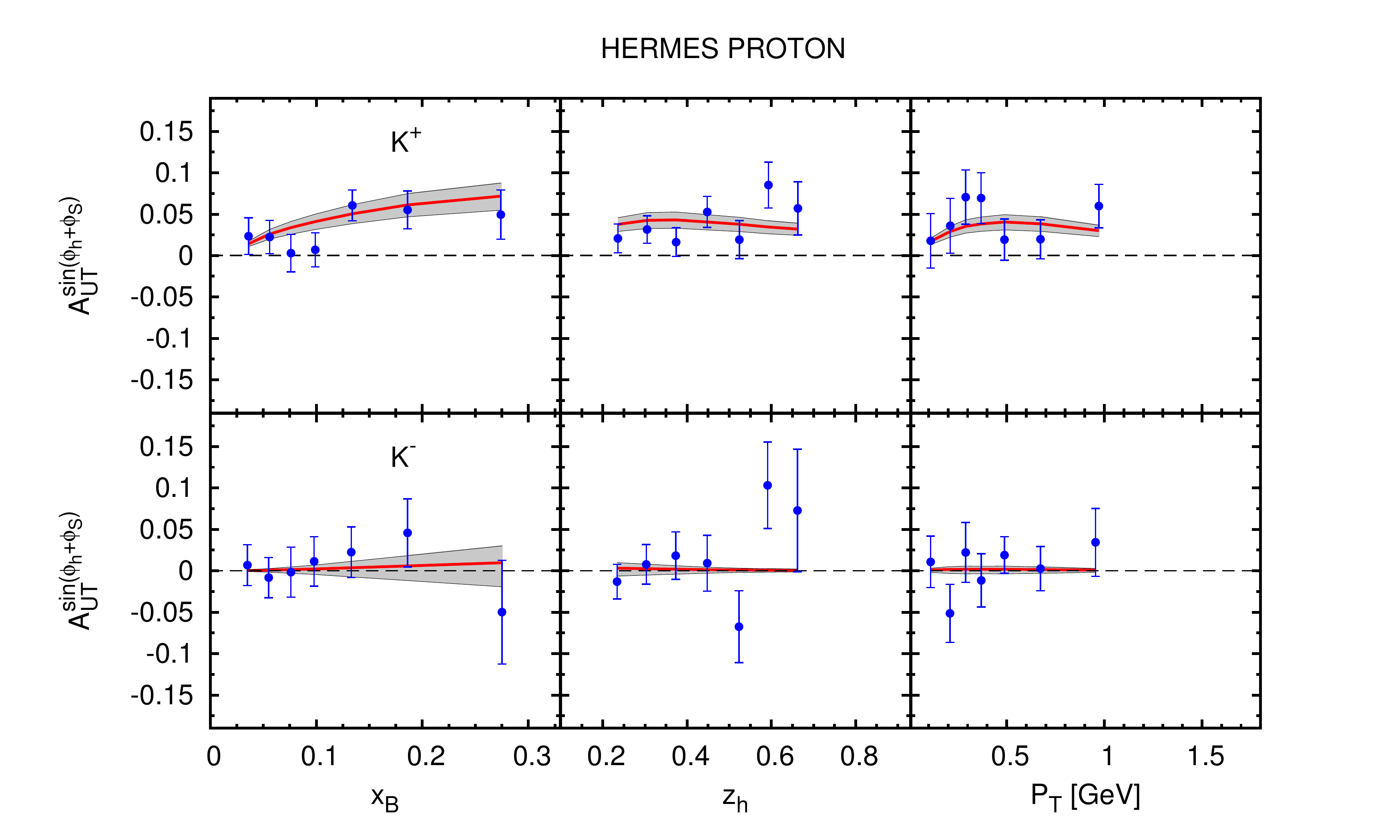}
\caption{The experimental data on the SIDIS azimuthal moment
$A_{UT}^{\sin(\phi_h+\phi_S)}$ as measured by the HERMES
Collaboration~\cite{Airapetian:2010ds}, are compared with our computation
of the same quantity. The solid (red) lines correspond to our reference fit, with the parameters given
in Eq.~\eqref{fitpar-K}. The shaded area corresponds to the statistical 
uncertainty on these parameters. For the transversity distributions we used 
the fixed parameters reported in Table~\ref{fitpar-pi}.}
\label{fig:an-hermes-k}
\end{figure}
%
\begin{figure}[ht]
\includegraphics[width=10.5truecm,angle=0]{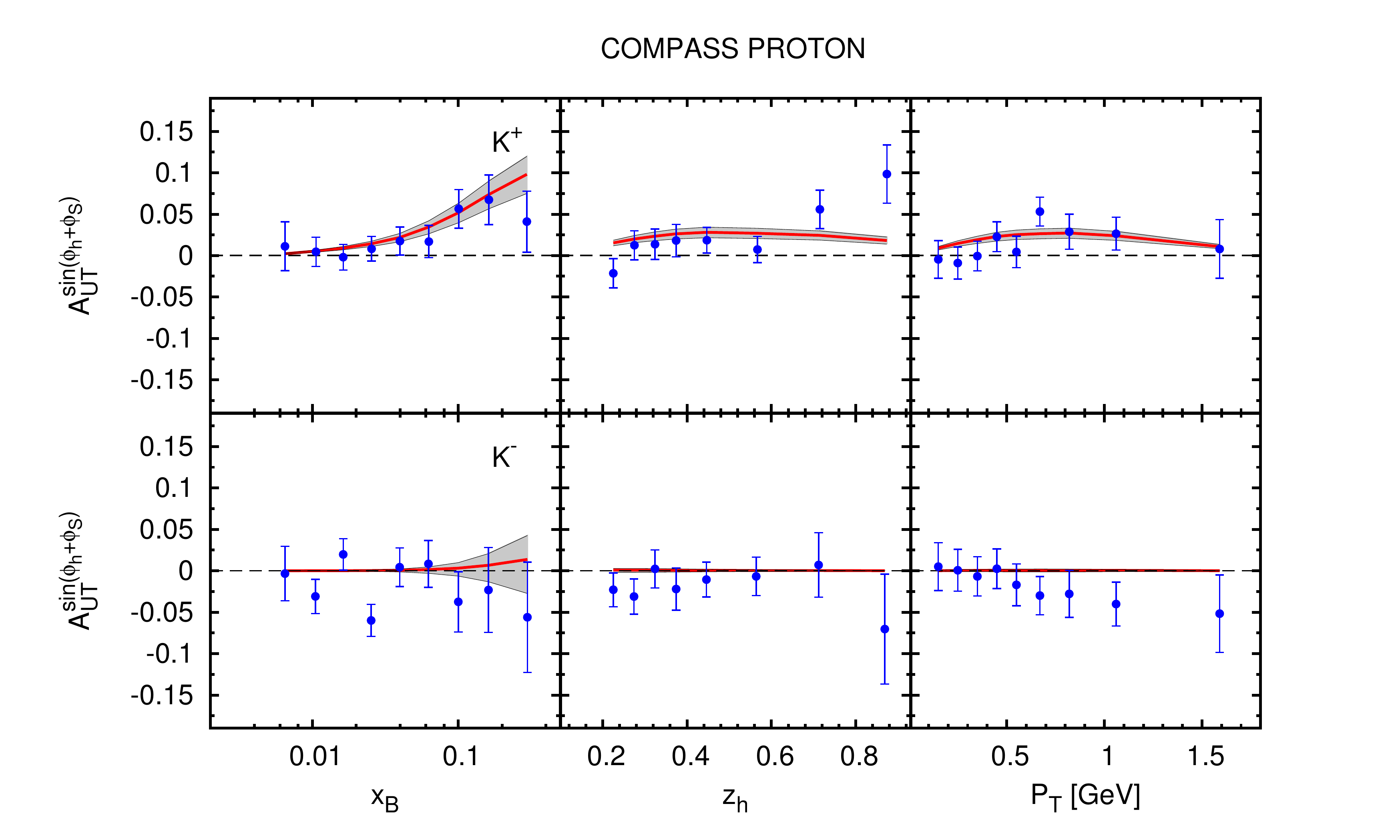}
\includegraphics[width=10.5truecm,angle=0]{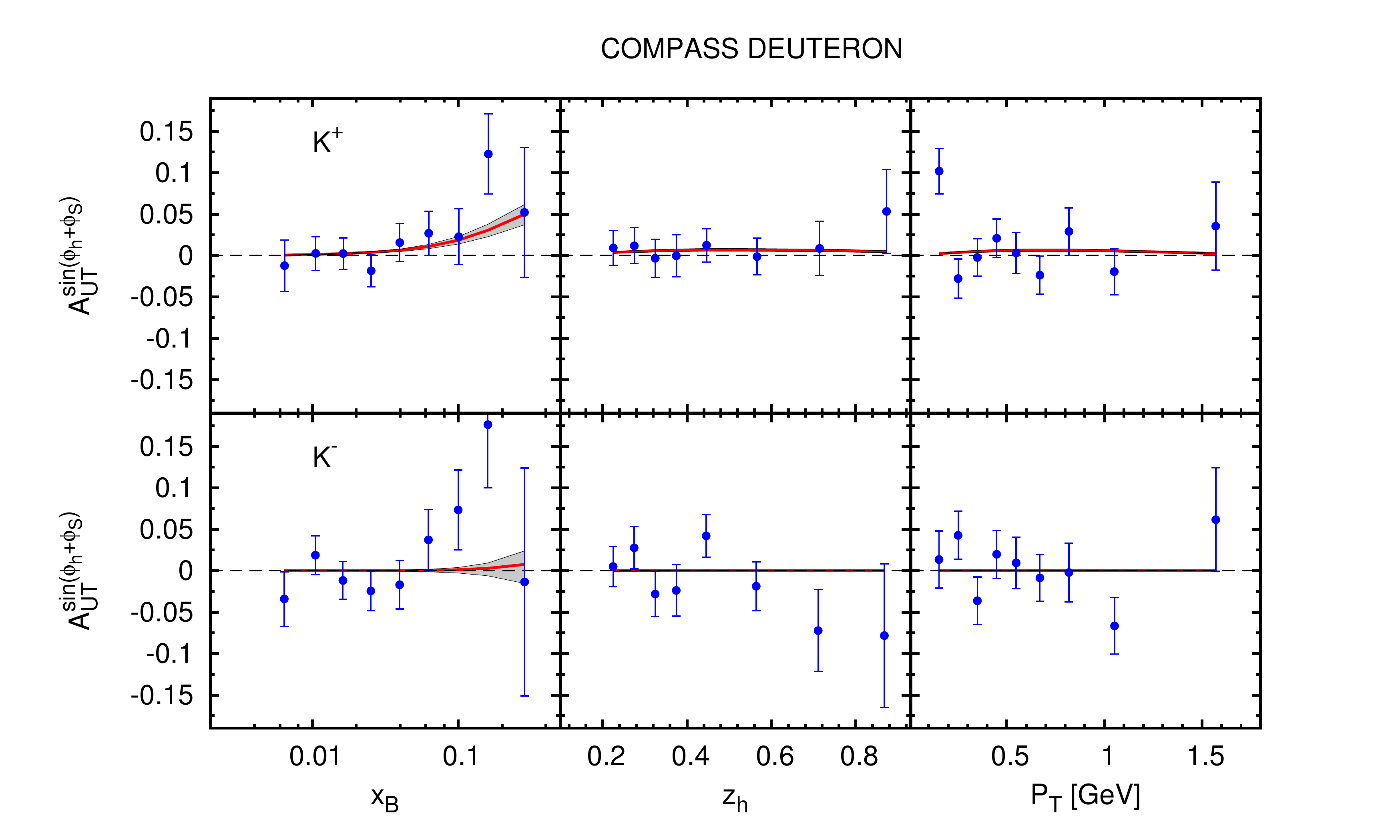}
\caption{The experimental data on the SIDIS 
Collins SSA $A_{UT}^{\sin(\phi_h+\phi_S)}$ as measured by the COMPASS Collaboration
on proton (upper panel)~\cite{Adolph:2014zba} and deuteron (lower panel) 
targets~\cite{Alekseev:2008aa},
are compared with our computation of the same quantity. 
The solid (red) lines correspond to our reference fit, with the parameters given
in Eq.~\eqref{fitpar-K}. The shaded area corresponds to the statistical 
uncertainty on these parameters. For the transversity distributions we used 
the fixed parameters reported in Table~\ref{fitpar-pi}.
}
\label{fig:an-compass-k}
\end{figure}
%
\end{enumerate}

\newpage

\subsection{Fits with additional parameters \label{more-param}}

Looking at the results of our reference fit, Eq.~\eqref{fitpar-K}, the
disfavoured Collins function appears to be quite undetermined and compatible
with zero, while the favoured one is definitely non-zero and positive.
However, we have 
assumed that the heavy ($s$ quark) and light ($u$ quark) 
favoured contributions are controlled by the same parameter. We wonder whether,
by disentangling these two contributions, one can confirm the results obtained
above.

An inspection of the analytical formulae,
Eqs.~\eqref{P0k}, \eqref{P0pk} and~\eqref{Dhh}, \eqref{Nhh}, 
shows that the sign of the light-flavour favoured contribution is determined by
the $\pi K$ data, where it appears convoluted with the pion Collins
function, which is fixed. Most of the information, in particular, comes
from the $A_0^{UL}$ asymmetries, which are dominated by doubly favoured
terms of the type $\tilde{\Delta} ^N D_{\pi^+/u^\ua} \tilde{\Delta}^N
D_{K^-/{\bar u}^\ua}$.

The heavy flavour  contribution, instead, is not determined by the data
(not even in sign): this is due to the fact that, in $K K$ production
processes, it appears in doubly favoured terms where it is convoluted
with itself and therefore insensitive to the sign choice, while in
$\pi K$ production processes it appears only in {\em sub-leading} combinations, such as $\tilde{\Delta} ^N 
D_{\pi^-/s^\ua} \tilde{\Delta}^N
D_{K^+/{\bar s}^\ua}$.

\begin{figure}[ht]
\begin{center}
\includegraphics[width=8.9cm,angle=0]{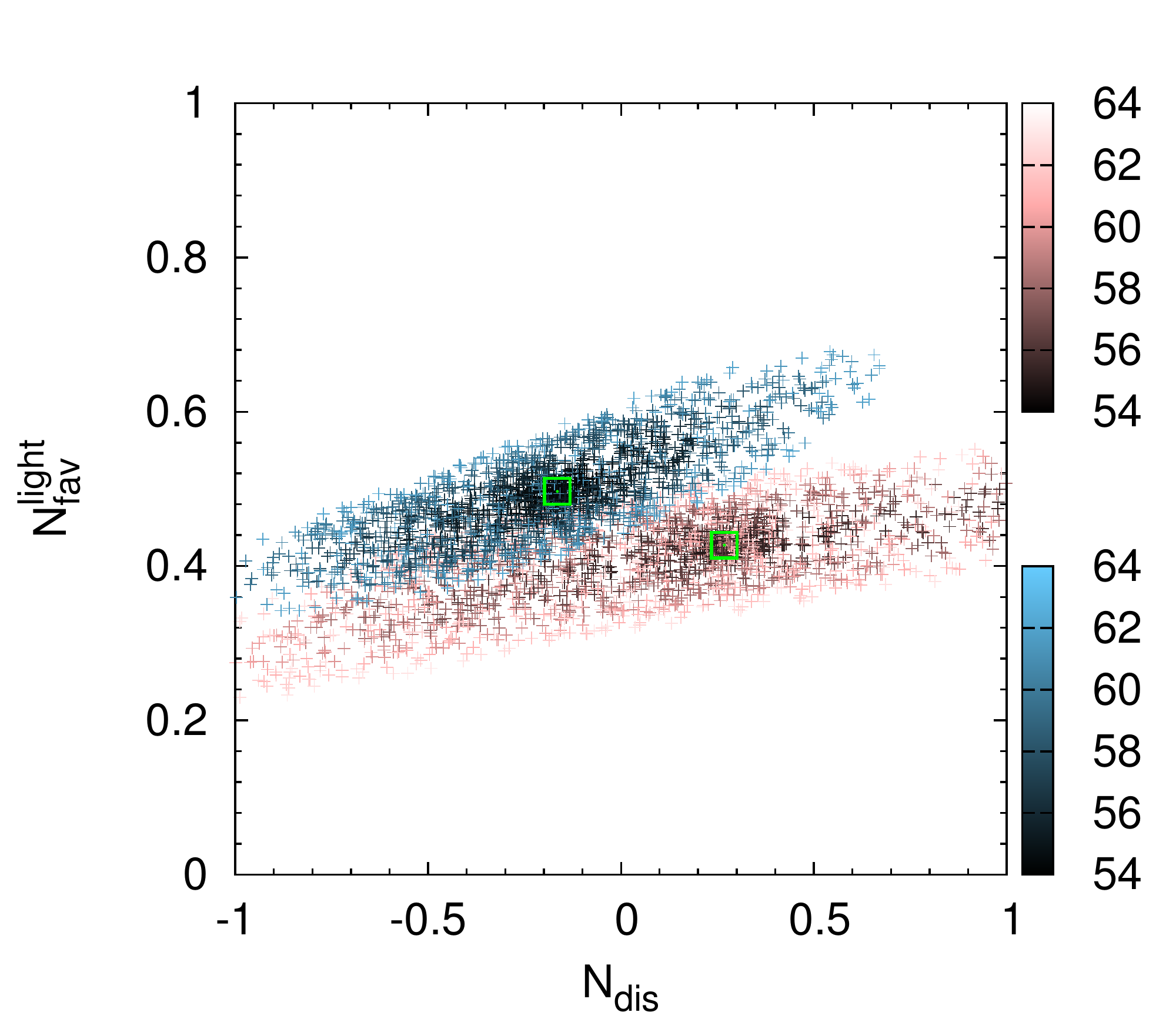}
\includegraphics[width=8.9cm,angle=0]{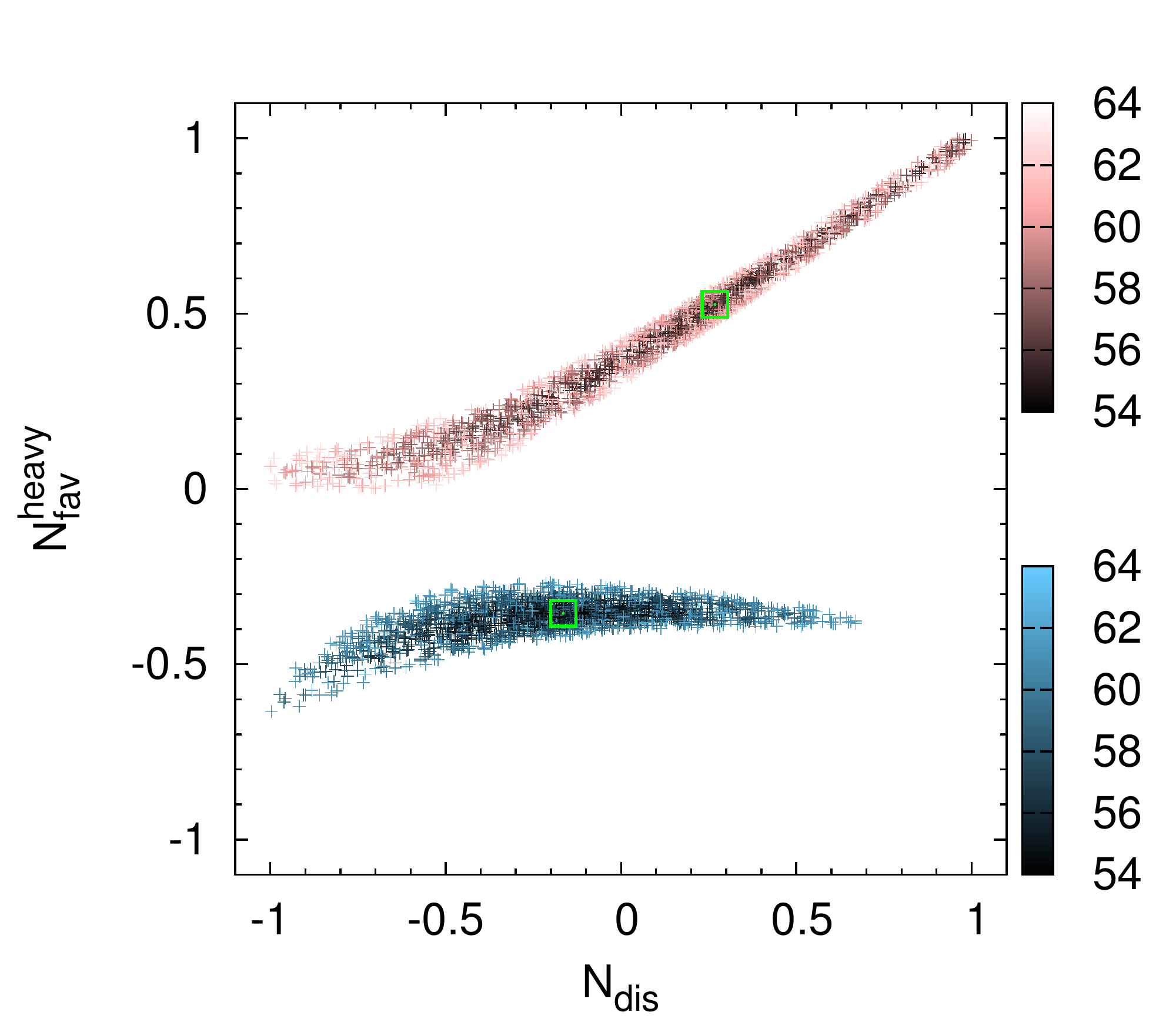}
\end{center}
\caption{Correlation between the parameters:  $N^{\rm light}_{\rm fav}$
and $N_{\rm dis}$ (left panel) and $N^{\rm heavy}_{\rm fav}$
and $N_{\rm dis}$ (right panel). Red points represent solutions with positive
$N^{\rm heavy}_{\rm fav}$, while blue points represent solutions
with negative $N^{\rm heavy}_{\rm fav}$. All points in the figure correspond
to a total $\chi^2$ included between $\chi^2_{\rm min}$ and
$\chi^2_{\rm min} + \Delta\chi^2$; the spread of the points
indicates the statistical error which affects the two parameters. 
Lighter(darker) shades of color represent higher(lower) values of $\chi^2$.
The points in which  $\chi^2 \equiv \chi^2_{\rm min}$ 
are shown as green squares.}
\label{fig:correlation}
\end{figure}

To study this in more detail, we have performed a series of fits
allowing for up to three free parameters, {\it i.e.} one normalisation
constant for the favoured light flavour, $N^{\rm light}_{\rm fav}$, one
for the favoured heavy flavour, $N^{\rm heavy}_{\rm fav}$, and one for the
disfavoured, $N_{\rm dis}$, contributions. The results, with the
$\chi^2_{\rm d.o.f.}$ for each of the fits, are presented in
Table~\ref{fit3}, while some correlations between the parameters are studied
in Fig.~\ref{fig:correlation}. Let us comment on such results.
\begin{itemize}
\item
The first clear conclusion is that it is not possible to fit the data with 
one and only one of the parameters $N^{\rm light}_{\rm fav}$, 
$N^{\rm heavy}_{\rm fav}$, $N_{\rm dis}$, as shown in the upper panel of 
Table~\ref{fit3}.
\item
Regarding the two parameter fits (central panel of Table~\ref{fit3}), we see that
the data can be successfully described only by including the light favoured
contribution together with either the heavy favoured or the disfavoured
Collins function. 
Notice that the sign of the heavy contribution, can be either positive or negative, leading to equally good 
fits (first two lines of the central panel in Table~\ref{fit3}). 
The sign of $N_{\rm fav}^{\rm light}$ turns out to be always positive, with its best value in the approximate range 
between $0.3$ and $0.6$ (see the left panel of Fig.~\ref{fig:correlation}).
Instead, fitting the data without any light quark favoured contribution appears
not to be possible (last two lines of the central panel in Table~\ref{fit3}).
\item
Fits with three parameters (bottom panel of Table~\ref{fit3}) result in good values of
$\chi^2_{\rm d.o.f.}$. These fits allow us to study the correlation among the
free parameters. We, in fact, observe a very strong correlation between the
heavy flavour (favoured) and the disfavoured contributions to the kaon
Collins functions: values of $N^{\rm heavy}_{\rm fav}$ with opposite sign can
easily be compensated by different values of $N_{\rm dis}$, resulting in fits
of equal quality, as shown in the last part of Table~\ref{fit3}. 
We actually find two distinct solutions resulting from the present data, 
one with positive and one with negative heavy flavour Collins FFs.

Fig.~\ref{fig:correlation} (right panel) illustrates this correlation. 
Two distinct distributions are clearly evident: 
red(blue) points
represent solutions with positive(negative) $N^{\rm heavy}_{\rm fav}$.
All points in the figure correspond to a total
$\chi^2$ included between $\chi^2_{\rm min}$ and
$\chi^2_{\rm min} + \Delta\chi^2$; for a three parameter fit
$\Delta\chi^2=8.02$. The spread of the points indicates the statistical
error which affects the two parameters.
Lighter(darker) shades of color represent higher(lower) values of $\chi^2$.
The points in which  $\chi^2 \equiv \chi^2_{\rm min}$ 
are shown as green squares.

Notice that model calculations predict the same sign of light  
and heavy flavour Collins FF, see for instance Ref.~\cite{Bacchetta:2007wc}.
\end{itemize}

\begin{figure}[ht]
\vspace*{-1cm}
\centering
\includegraphics[width=8.0truecm,angle=0]{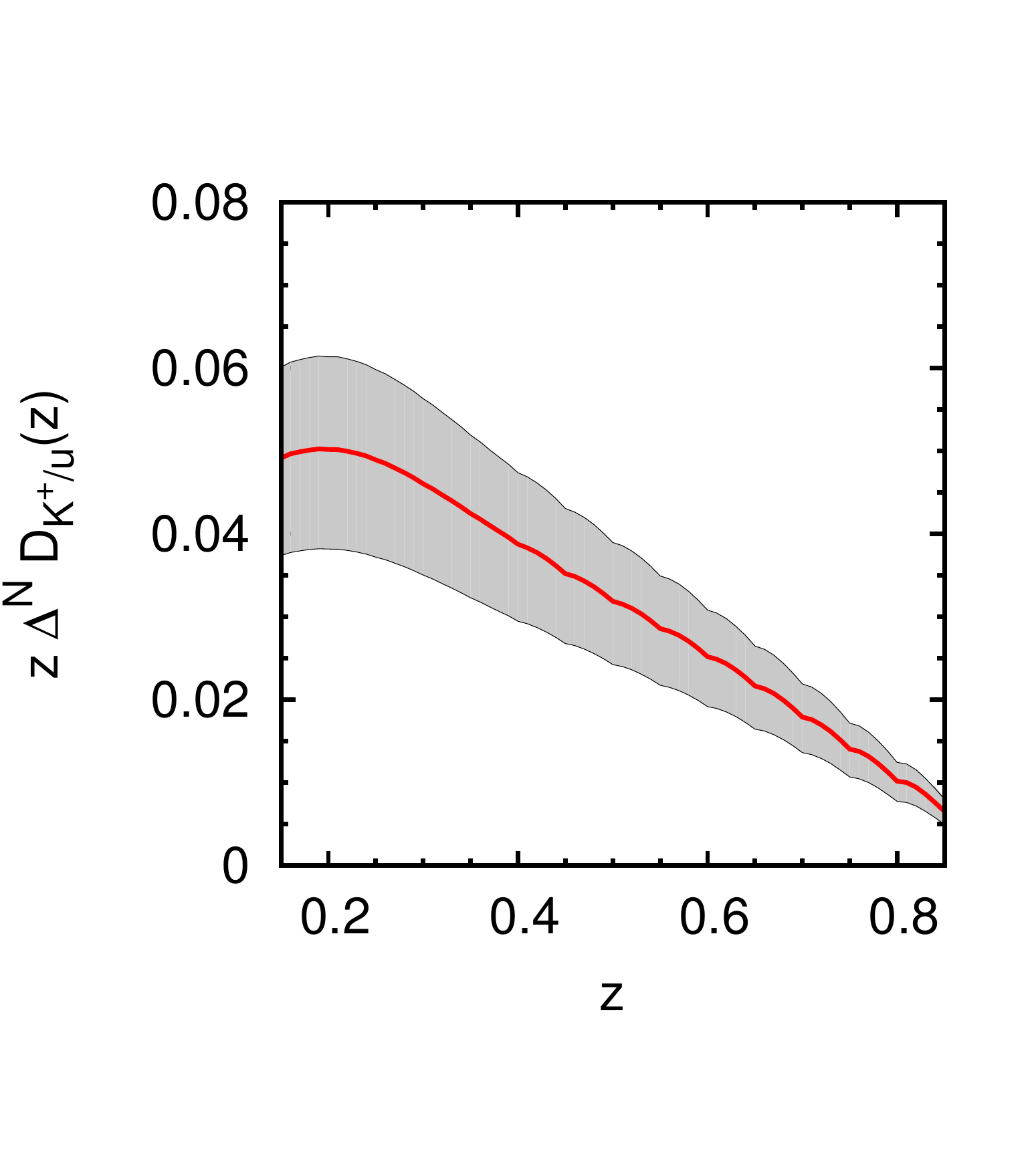}
\vspace*{-1cm}
\caption{Plot of $z$ times the lowest $\bfp_\perp$-moment, Eqs.~\eqref{p-0mom-D} 
and~\eqref{D-mom-tilde}, of the $u^\uparrow \to K^+ \,X$ Collins function, as 
extracted in our reference fit (with the parameters of Eq.~\eqref{fitpar-K}). 
The analogous plots for heavy flavour favoured  and (all flavour) disfavoured 
Collins functions are not shown: in fact, it is not possible to reliably 
distinguish between these two contributions to the available BaBar data. 
Furthermore, not even the sign of the heavy flavour favoured Collins function 
can be determined.
}
\label{fig:collins-K}
\end{figure}

In Fig.~\ref{fig:collins-K} we show the lowest $\bfp_\perp$-moment of the 
light-flavour favoured kaon Collins function, as extracted in our reference 
fit (with the parameters of Eq.~\eqref{fitpar-K}).
Note that, in the case of a factorised Gaussian shape,
Eqs.~(\ref{coll-funct}),~(\ref{coll-D}) and (\ref{hpcollins}),
the lowest $\bfp_\perp$-moment of the Collins function,
\begin{equation}
\Delta^N D_{h/q^\uparrow}(z,Q^2) = \int\,d^2\bfp_\perp\,
\Delta^N D_{h/q^\uparrow}(z,p_\perp,Q^2)\,,
\label{p-0mom-D}
\end{equation}
%
%
%
%
is related to the $z$-dependent part of the Collins function,
$\tilde{\Delta}^N D_{h/q^\uparrow}(z,Q^2)$, by 
\be
\Delta ^N D_{h/q^\ua}(z, Q^2) = \frac{\sqrt{\pi}}{2} \,
\frac{\avp_{_{\!C}}^{3/2}}{\avp} \, \frac{\sqrt{2e}}{M_C} \,
\tilde{\Delta} ^N D_{h/q^\ua}(z, Q^2)\,.
\label{D-mom-tilde}
\ee

The heavy flavour favoured and (all flavour) disfavoured results are not 
shown: in fact, the study performed above shows that it is not possible to 
reliably distinguish between these two contributions to the available data. 
Furthermore, not even the sign of the heavy flavour favoured Collins function 
can be determined.

%
\begin{table}[ht]
\renewcommand{\tabcolsep}{0.4pc} 
\renewcommand{\arraystretch}{1.5} 
\begin{tabular}{|ccccc|}
\hline
 ~ $N_{\rm fav}^{\rm light}$ ~    & ~ $N_{\rm fav}^{\rm heavy} > 0$ ~ &
 ~ $N_{\rm fav}^{\rm heavy}<0$ ~  & ~ $N_{\rm dis}$ & ~ $\chi^2_{\rm d.o.f.}$ ~ \\
\hline
\hline
\yesp &   \nop & \nop    &  \nop  & 1.83 \\
\nop   &   \yesp & \nop
  &  \nop  & 3.32 \\
\nop   &   \nop
 & \yesp  &  \nop  & 5.68 \\
\nop   &   \nop  & \nop  &  \yesp  & 3.94 \\
\hline
\hline
\yesp   &  \yesp & \nop
     & \nop  & 0.89 \\
\yesp   &  \nop
   & \yesp     &  \nop  & 0.88 \\
\yesp   &   \nop & \nop    &   \yesp & 0.98 \\
\nop   &   \yesp & \nop
       &  \yesp  & 2.00\\
\nop   &   \nop
 & \yesp       &  \yesp  & 4.00\\
\hline
\hline
\yesp   &  \yesp & \nop
    &   \yesp & 0.90 \\
\yesp   &  \nop
 & \yesp    &   \yesp & 0.89 \\
\hline
\end{tabular}
\caption{$\chi^2$/d.o.f. for different scenarios for the kaon Collins functions: one-parameter (upper panel), 
two-parameter (central panel)  and three-parameter (lower panel) fits. The symbol {\color{red}\textbullet}~means that 
the corresponding
parameter is actually used in the fit, while the symbol
{\color{blue}\textopenbullet}~means that the contribution to the Collins asymmetry corresponding to that parameter is
not included in the fit. For $N_{\rm fav}^{\rm heavy}$, we explicitly indicate the two different constraints we use:  
$N_{\rm fav}^{\rm heavy} > 0$ and $N_{\rm fav}^{\rm heavy}<0$.
\label{fit3}}
\end{table}
%
%
%

%
%

\section{Comments and conclusions}\label{Com}

We have extracted, for the first time, the kaon Collins functions,
$\qup \to K\,X$, by best fitting recent BaBar data~\cite{Aubert:2015hha}.
This paper extends a recent study of the Collins functions in $e^+e^-$
and SIDIS processes~\cite{Anselmino:2015sxa} limited to pion production.

It turns out that a simple phenomenological parameterisation
of the Collins function, Eqs.~\eqref{coll-funct} and~\eqref{coll-D}, is
quite adequate to describe the data. When comparing with the pion Collins
functions~\cite{Anselmino:2015sxa}, 
due to the limited amount and relatively big errors of data,
an even smaller number of parameters suffices to describe the experimental
results. Indeed, we find that kaon Collins functions of two kinds, favoured
and disfavoured, both simply proportional to the 
unpolarised TMD fragmentation functions, describe well the BaBar data.

As a result of the attempted fits, we can conclude that a definite
outcome of this study is the determination of a {\em positive} 
$u^\uparrow \to K^+ X =
\bar u^\uparrow \to K^- X$ Collins function, assuming a {\em positive}
favoured pion Collins function \cite{Anselmino:2015sxa}. No definite independent
conclusion, based on the available data, can be drawn on the signs of $s^\uparrow
\to K^- X = \bar s^\uparrow \to K^+ X$ Collins functions and on the
disfavoured ones. 

The extracted kaon Collins functions, together with the transversity distributions 
obtained in Ref.~\cite{Anselmino:2015sxa}, give a very good description, within 
the rather large experimental 
uncertainties, of SIDIS data on kaon Collins asymmetries measured by 
COMPASS~\cite{Adolph:2014zba,Alekseev:2008aa} and 
HERMES~\cite{Airapetian:2010ds} Collaborations.
This points towards a consistent and universal role of the Collins effect
in different physical processes, which should be further explored in the
future.

\acknowledgments
\noindent
M.A., M.B., J.O.G.H.~and S.M.~acknowledge the support of ``Progetto di Ricerca Ateneo/CSP" (codice
TO-Call3-2012-0103).

\bibliography{sample}

\begin{thebibliography}{17}
\expandafter\ifx\csname natexlab\endcsname\relax\def\natexlab#1{#1}\fi
\expandafter\ifx\csname bibnamefont\endcsname\relax
  \def\bibnamefont#1{#1}\fi
\expandafter\ifx\csname bibfnamefont\endcsname\relax
  \def\bibfnamefont#1{#1}\fi
\expandafter\ifx\csname citenamefont\endcsname\relax
  \def\citenamefont#1{#1}\fi
\expandafter\ifx\csname url\endcsname\relax
  \def\url#1{\texttt{#1}}\fi
\expandafter\ifx\csname urlprefix\endcsname\relax\def\urlprefix{URL }\fi
\providecommand{\bibinfo}[2]{#2}
\providecommand{\eprint}[2][]{\url{#2}}

\bibitem[{\citenamefont{Anselmino et~al.}(2007)\citenamefont{Anselmino,
  Boglione, D'Alesio, Kotzinian, Murgia et~al.}}]{Anselmino:2007fs}
\bibinfo{author}{\bibfnamefont{M.}~\bibnamefont{Anselmino}},
  \bibinfo{author}{\bibfnamefont{M.}~\bibnamefont{Boglione}},
  \bibinfo{author}{\bibfnamefont{U.}~\bibnamefont{D'Alesio}},
  \bibinfo{author}{\bibfnamefont{A.}~\bibnamefont{Kotzinian}},
  \bibinfo{author}{\bibfnamefont{F.}~\bibnamefont{Murgia}},
  \bibnamefont{et~al.}, \bibinfo{journal}{Phys. Rev.}
  \textbf{\bibinfo{volume}{D75}}, \bibinfo{pages}{054032}
  (\bibinfo{year}{2007}), \eprint{hep-ph/0701006}.

\bibitem[{\citenamefont{Anselmino et~al.}(2009)\citenamefont{Anselmino,
  Boglione, D'Alesio, Kotzinian, Murgia et~al.}}]{Anselmino:2008jk}
\bibinfo{author}{\bibfnamefont{M.}~\bibnamefont{Anselmino}},
  \bibinfo{author}{\bibfnamefont{M.}~\bibnamefont{Boglione}},
  \bibinfo{author}{\bibfnamefont{U.}~\bibnamefont{D'Alesio}},
  \bibinfo{author}{\bibfnamefont{A.}~\bibnamefont{Kotzinian}},
  \bibinfo{author}{\bibfnamefont{F.}~\bibnamefont{Murgia}},
  \bibnamefont{et~al.}, \bibinfo{journal}{Nucl. Phys. Proc. Suppl.}
  \textbf{\bibinfo{volume}{191}}, \bibinfo{pages}{98} (\bibinfo{year}{2009}),
  \eprint{0812.4366}.

\bibitem[{\citenamefont{Anselmino et~al.}(2013)\citenamefont{Anselmino,
  Boglione, D'Alesio, Melis, Murgia et~al.}}]{Anselmino:2013vqa}
\bibinfo{author}{\bibfnamefont{M.}~\bibnamefont{Anselmino}},
  \bibinfo{author}{\bibfnamefont{M.}~\bibnamefont{Boglione}},
  \bibinfo{author}{\bibfnamefont{U.}~\bibnamefont{D'Alesio}},
  \bibinfo{author}{\bibfnamefont{S.}~\bibnamefont{Melis}},
  \bibinfo{author}{\bibfnamefont{F.}~\bibnamefont{Murgia}},
  \bibnamefont{et~al.}, \bibinfo{journal}{Phys. Rev.}
  \textbf{\bibinfo{volume}{D87}}, \bibinfo{pages}{094019}
  (\bibinfo{year}{2013}), \eprint{1303.3822}.

\bibitem[{\citenamefont{Anselmino et~al.}(2015)\citenamefont{Anselmino,
  Boglione, D'Alesio, Hernandez, Melis, Murgia, and
  Prokudin}}]{Anselmino:2015sxa}
\bibinfo{author}{\bibfnamefont{M.}~\bibnamefont{Anselmino}},
  \bibinfo{author}{\bibfnamefont{M.}~\bibnamefont{Boglione}},
  \bibinfo{author}{\bibfnamefont{U.}~\bibnamefont{D'Alesio}},
  \bibinfo{author}{\bibfnamefont{J.~O.~G.} \bibnamefont{Hernandez}},
  \bibinfo{author}{\bibfnamefont{S.}~\bibnamefont{Melis}},
  \bibinfo{author}{\bibfnamefont{F.}~\bibnamefont{Murgia}}, \bibnamefont{and}
  \bibinfo{author}{\bibfnamefont{A.}~\bibnamefont{Prokudin}}
  (\bibinfo{year}{2015}), \eprint{1510.05389}.

\bibitem[{\citenamefont{Airapetian et~al.}(2010)}]{Airapetian:2010ds}
\bibinfo{author}{\bibfnamefont{A.}~\bibnamefont{Airapetian}}
  \bibnamefont{et~al.} (\bibinfo{collaboration}{HERMES Collaboration}),
  \bibinfo{journal}{Phys. Lett.} \textbf{\bibinfo{volume}{B693}},
  \bibinfo{pages}{11} (\bibinfo{year}{2010}), \eprint{1006.4221}.

\bibitem[{\citenamefont{Airapetian et~al.}(2013)}]{Airapetian:2012yg}
\bibinfo{author}{\bibfnamefont{A.}~\bibnamefont{Airapetian}}
  \bibnamefont{et~al.} (\bibinfo{collaboration}{HERMES Collaboration}),
  \bibinfo{journal}{Phys. Rev.} \textbf{\bibinfo{volume}{D87}},
  \bibinfo{pages}{012010} (\bibinfo{year}{2013}), \eprint{1204.4161}.

\bibitem[{\citenamefont{Alekseev et~al.}(2009)}]{Alekseev:2008aa}
\bibinfo{author}{\bibfnamefont{M.}~\bibnamefont{Alekseev}} \bibnamefont{et~al.}
  (\bibinfo{collaboration}{COMPASS Collaboration}), \bibinfo{journal}{Phys.
  Lett.} \textbf{\bibinfo{volume}{B673}}, \bibinfo{pages}{127}
  (\bibinfo{year}{2009}), \eprint{0802.2160}.

\bibitem[{\citenamefont{Adolph et~al.}(2015)}]{Adolph:2014zba}
\bibinfo{author}{\bibfnamefont{C.}~\bibnamefont{Adolph}} \bibnamefont{et~al.}
  (\bibinfo{collaboration}{COMPASS Collaboration}), \bibinfo{journal}{Phys.
  Lett.} \textbf{\bibinfo{volume}{B744}}, \bibinfo{pages}{250}
  (\bibinfo{year}{2015}), \eprint{1408.4405}.

\bibitem[{\citenamefont{Aubert et~al.}(2015)}]{Aubert:2015hha}
\bibinfo{author}{\bibfnamefont{B.}~\bibnamefont{Aubert}} \bibnamefont{et~al.}
  (\bibinfo{collaboration}{BaBar Collaboration}) (\bibinfo{year}{2015}),
  \eprint{1506.05864}.

\bibitem[{\citenamefont{Collins and Metz}(2004)}]{Collins:2004nx}
\bibinfo{author}{\bibfnamefont{J.~C.} \bibnamefont{Collins}} \bibnamefont{and}
  \bibinfo{author}{\bibfnamefont{A.}~\bibnamefont{Metz}},
  \bibinfo{journal}{Phys. Rev. Lett.} \textbf{\bibinfo{volume}{93}},
  \bibinfo{pages}{252001} (\bibinfo{year}{2004}), \eprint{hep-ph/0408249}.

\bibitem[{\citenamefont{Anselmino et~al.}(2014)\citenamefont{Anselmino,
  Boglione, Gonzalez~Hernandez, Melis, and Prokudin}}]{Anselmino:2013lza}
\bibinfo{author}{\bibfnamefont{M.}~\bibnamefont{Anselmino}},
  \bibinfo{author}{\bibfnamefont{M.}~\bibnamefont{Boglione}},
  \bibinfo{author}{\bibfnamefont{J.}~\bibnamefont{Gonzalez~Hernandez}},
  \bibinfo{author}{\bibfnamefont{S.}~\bibnamefont{Melis}}, \bibnamefont{and}
  \bibinfo{author}{\bibfnamefont{A.}~\bibnamefont{Prokudin}},
  \bibinfo{journal}{JHEP} \textbf{\bibinfo{volume}{1404}}, \bibinfo{pages}{005}
  (\bibinfo{year}{2014}), \eprint{1312.6261}.

\bibitem[{\citenamefont{Gluck et~al.}(1998)\citenamefont{Gluck, Reya, and
  Vogt}}]{Gluck:1998xa}
\bibinfo{author}{\bibfnamefont{M.}~\bibnamefont{Gluck}},
  \bibinfo{author}{\bibfnamefont{E.}~\bibnamefont{Reya}}, \bibnamefont{and}
  \bibinfo{author}{\bibfnamefont{A.}~\bibnamefont{Vogt}},
  \bibinfo{journal}{Eur.Phys.J.} \textbf{\bibinfo{volume}{C5}},
  \bibinfo{pages}{461} (\bibinfo{year}{1998}), \eprint{hep-ph/9806404}.

\bibitem[{\citenamefont{de~Florian et~al.}(2007)\citenamefont{de~Florian,
  Sassot, and Stratmann}}]{deFlorian:2007hc}
\bibinfo{author}{\bibfnamefont{D.}~\bibnamefont{de~Florian}},
  \bibinfo{author}{\bibfnamefont{R.}~\bibnamefont{Sassot}}, \bibnamefont{and}
  \bibinfo{author}{\bibfnamefont{M.}~\bibnamefont{Stratmann}},
  \bibinfo{journal}{Phys.Rev.} \textbf{\bibinfo{volume}{D76}},
  \bibinfo{pages}{074033} (\bibinfo{year}{2007}), \eprint{0707.1506}.

\bibitem[{\citenamefont{Lees et~al.}(2014)}]{TheBABAR:2013yha}
\bibinfo{author}{\bibfnamefont{J.}~\bibnamefont{Lees}} \bibnamefont{et~al.}
  (\bibinfo{collaboration}{BaBar Collaboration}), \bibinfo{journal}{Phys.Rev.}
  \textbf{\bibinfo{volume}{D90}}, \bibinfo{pages}{052003}
  (\bibinfo{year}{2014}), \eprint{1309.5278}.

\bibitem[{\citenamefont{Seidl et~al.}(2008)}]{Seidl:2008xc}
\bibinfo{author}{\bibfnamefont{R.}~\bibnamefont{Seidl}} \bibnamefont{et~al.}
  (\bibinfo{collaboration}{Belle Collaboration}), \bibinfo{journal}{Phys. Rev.}
  \textbf{\bibinfo{volume}{D78}}, \bibinfo{pages}{032011}
  (\bibinfo{year}{2008}), \eprint{0805.2975}.

\bibitem[{\citenamefont{Seidl et~al.}(2012)}]{Seidl:2012er}
\bibinfo{author}{\bibfnamefont{R.}~\bibnamefont{Seidl}} \bibnamefont{et~al.}
  (\bibinfo{collaboration}{Belle Collaboration}), \bibinfo{journal}{Phys. Rev.}
  \textbf{\bibinfo{volume}{D86}}, \bibinfo{pages}{032011(E)}
  (\bibinfo{year}{2012}), \eprint{0805.2975}.

\bibitem[{\citenamefont{Bacchetta et~al.}(2008)\citenamefont{Bacchetta,
  Gamberg, Goldstein, and Mukherjee}}]{Bacchetta:2007wc}
\bibinfo{author}{\bibfnamefont{A.}~\bibnamefont{Bacchetta}},
  \bibinfo{author}{\bibfnamefont{L.~P.} \bibnamefont{Gamberg}},
  \bibinfo{author}{\bibfnamefont{G.~R.} \bibnamefont{Goldstein}},
  \bibnamefont{and}
  \bibinfo{author}{\bibfnamefont{A.}~\bibnamefont{Mukherjee}},
  \bibinfo{journal}{Phys. Lett.} \textbf{\bibinfo{volume}{B659}},
  \bibinfo{pages}{234} (\bibinfo{year}{2008}), \eprint{0707.3372}.

\end{thebibliography}

\end{document}